\documentclass[11pt,a4paper]{article}
 \usepackage[dvips]{graphics}
 \usepackage{graphicx}
 \usepackage{afterpage}
 \usepackage{enumerate}
  \usepackage{longtable}

\newcount\longrefs
\def\aap{\ifnum\longrefs=1 {Astron.\ Astrophys.}\else
{A\hbox{\rm \&}A}\fi}
\def\aapr{\ifnum\longrefs=1 {Astron.\ Astrophys.\ Rev.}\else
{A\hbox{\rm \&}AR}\fi}
\def\aaps{\ifnum\longrefs=1 {Astron.\ Astrophys.\ Suppl.}\else
{A\hbox{\rm \&}A Suppl.}\fi}
\def\aj{\ifnum\longrefs=1 {Astron.\ J.}\else
{AJ}\fi}
\def\ao{\ifnum\longrefs=1 {Applied Optics}\else
{Appl.\ Opt.}\fi}
\def\aspcs{\ifnum\longrefs=1 {Astron.\ Soc.\ Pacific Conf. Series}\else
{ASP Conf.\ Ser.}\fi}
\def\apj{\ifnum\longrefs=1 {Astrophys.\ J.}\else
{ApJ}\fi}
\def\apjl{\ifnum\longrefs=1 {Astrophys.\ J. Lett.}\else
{ApJ}\fi}
\def\aplett{\ifnum\longrefs=1 {Astrophys.\ J. Lett.}\else
{ApJ}\fi}
\def\apjs{\ifnum\longrefs=1 {Astrophys.\ J. Suppl.}\else
{ApJS}\fi}
\def\apss{\ifnum\longrefs=1 {Astrophys.\ and Space Science}\else
{Astrophys.\ Space Sci.}\fi}
\def\araa{\ifnum\longrefs=1 {Ann.\ Rev.\ Astron.\ Astrophys.}\else
{ARA\hbox{\rm \&}A}\fi}
\def\azh{\ifnum\longrefs=1 {Astronomicheskii Zhurnal}\else
{Astron.\ Zhur.}\fi}
\def\baas{\ifnum\longrefs=1 {Bull.\ Am.\ Astron.\ Soc.}\else
{BAAS}\fi}
\def\bain{\ifnum\longrefs=1 {Bull.\ Astronom.\ Institutes Netherlands}\else
{Bull.\ Astr.\ Inst.\ Neth.}\fi}
\def\gca{\ifnum\longrefs=1 {Geochim.\ Cosmochim.\ Acta}\else
{Geochim.\ Cosmochim.\ Acta}\fi}
\def\grl{\ifnum\longrefs=1 {Geophys.\ Res.\ Lett.}\else
{Geoph.\ Res.\ Lett.}\fi}
\def\iaucirc{\ifnum\longrefs=1 {IAU Circulars}\else
{IAU Circ.}\fi}
\def\ip{\ifnum\longrefs=1 {in press}\else
{in press}\fi}
\def\jgr{\ifnum\longrefs=1 {J.\ Geophys.\ Res.}\else
{J.\ Geophys.\ Res.}\fi}
\def\jrasc{\ifnum\longrefs=1 {J.\ Royal Astron.\ Soc.\ Canada}\else
{JRAS Can.}\fi}
\def\mnras{\ifnum\longrefs=1 {Mon.\ Not.\ Roy.\ Astron.\ Soc.}\else
{MNRAS}\fi}
\def\nat{\ifnum\longrefs=1 {Nature}\else
{Nat}\fi}
\def\pasj{\ifnum\longrefs=1 {Pub.\ Astron.\ Soc.\ Japan}\else
{PASJ}\fi}
\def\pasp{\ifnum\longrefs=1 {Pub.\ Astron.\ Soc.\ Pacific}\else
{PASP}\fi}
\def\physscr{\ifnum\longrefs=1 {Physica Scripta}\else
{Phys.\ Scrip.}\fi}
\def\planss{\ifnum\longrefs=1 {Planetary \& Space Science}\else
{Plan. \& Space Sci.}\fi}
\def\procspie{\ifnum\longrefs=1 {Proc.\ SPIE}\else
{Proc.\ SPIE}\fi}
\def\qjras{\ifnum\longrefs=1 {Quarterly J.\ Royal Astron.\ Soc.}\else
{QJRAS}\fi}
\def\sa{\ifnum\longrefs=1 {Soviet Astron..}\else
{Sov.\ Astron.}\fi}
\def\skytel{\ifnum\longrefs=1 {Sky \& Telescope}\else
{Sky \& Tel.}\fi}
\def\solphys{\ifnum\longrefs=1 {Solar Phys.}\else
{Sol.\ Phys.}\fi}
\def\ssr{\ifnum\longrefs=1 {Space Science Rev.}\else
{Space\ Sci.\ Rev.}\fi}

\begin{document}
%%==========================================================

 \title{\bf Fourier Analysis of Spectra of Solar-Type Stars}
\author{\bf V. A. Sheminova}
 \date{}

 \maketitle
 \thanks{}
\begin{center}
{Main Astronomical Observatory, National Academy of Sciences of
Ukraine,
\\Akademika  Zabolotnoho 27,  Kyiv,  03143 Ukraine\\ e-mail: shem@mao.kiev.ua}
\end{center}

 \begin{abstract}

{We used Fourier transform techniques to identify macroturbulent velocity. The analysis is done with mictoturbulent velocity and  rotation velocity as an unknown  quantities. In order to distinguish the effects of rotation from macroturbulence effects in slowly rotating stars, primarily the main lobe of residual Fourier transforms of the observed lines, which were taken from the solar spectrum and the spectra of two other stars, was used. This case of Fourier analysis of spectral lines is the most complicated one. The end results were in a satisfactory agreement with the data obtained using different methods. The average values of microturbulent, macroturbulent, and rotation velocities were 0.85, 2.22, and 1.75 km/s for the Sun as the star; 0.58, 1.73, and 0.78 km/s for HD~10700; and 1.16, 3.56, and 6.24 km/s for HD 1835. It was found that the macroturbulent velocity decreases with height in the atmosphere of the Sun and HD~1835. In the case of HD 10700, the macroturbulent velocity did not change with height, and the determined rotation velocity was two times lower than the one obtained using other methods. It was concluded that Fourier transform techniques are suitable for determining the velocities in atmospheres of solar-type stars with very slow rotation.
}
\end{abstract}
%-------------------------------------------------

\section{Introduction }

Modern theoretical three-dimensional (3D) hydrodynamic atmosphere models, which describe a selfconsistent velocity field, are lacking for many stars. Therefore, one-dimensional (1D) models are often used to synthesize spectra. The micro- and macroturbulence approximation is normally adopted in order to describe the velocity field in these 1D models. The determination of micro- and macroturbulence parameters based on the observed stellar spectra is a relevant problem. These parameters are required (especially when the equivalent width of a spectral line is not known, and one is forced to use the spectral line profile) to determine the chemical composition of stars with 1D models. They are also needed to test theoretical 3Datmosphere models. The micro- and macroturbulence parameters for the Sun and for many other stars are updated constantly. The microturbulent velocity is much easier to determine. Various methods utilizing the growth curves, FWHM values, profiles, and equivalent widths of spectral lines may be used for this purpose. Only the line profiles may be used to determine the macroturbulent velocity, and then only on the condition that the stellar rotation velocity is known. The difficulty lies in the fact that macroturbulence and rotation have almost the same effect on the line profile. The Fourier method is used in order to distinguish
between these effects: the Fourier transform of spectral lines is analyzed instead of the line profiles.

The Fourier method for analysis of stellar spectra is well developed \cite{1971A&A....13..169B, 1976oasp.book.....G, 1977ApJ...218..530G, 1982ApJ...262..682G, 2014AJ....147...81G, 2006PASP..118.1112G, 1976ApJ...208..487S}, but it is rarely used in current studies that rely on automated calculations for large stellar samples  (e.g., \cite{2005ApJS..159..141V}). Line blending, low spectral resolution and signal-to-noise ratio, and the cases of very slow stellar rotation present considerable difficulties for the Fourier method.

The aim of this study was to demonstrate the applicability of the Fourier method in the evaluation of parameters of microturbulent, macroturbulent, and rotation velocities of solar-type stars with very slow rotation. The solar spectrum lines and lines in the spectrum of two solar-type stars (HD~1835 and HD~10700) were used to test the Fourier technique.

\section{Fourier method for stellar spectra}

The Fourier method and its application to stellar spectra were described in most detail by Gray   \cite{1973ApJ...184..461G, 1975ApJ...202..148G, 1976oasp.book.....G, 1977ApJ...218..530G, 1982ApJ...262..682G}. The method is based on the convolution theorem, which states that the Fourier transform of a convolution of two functions is the product of their Fourier transforms. Since this theorem is applicable only to independent functions, it requires a number of approximations to use Fourier analysis to determine the velocities of nonthermal motion in stellar atmospheres. First, it is assumed that nonthermal kinematic line broadening is induced by microturbulence with isotropic Gaussian distribution, macroturbulence with isotropic Gaussian distribution, and solid-body rotation of a star.

According to \cite{1973ApJ...184..461G, 1977ApJ...218..530G} the observed line profile for flux $D(\lambda)$ may be presented as a multiple convolution of the true line flux  $F_{\lambda}^0$  profile without macroturbulent and rotational broadening, instrumental profile $I(\lambda)$, macroturbulent broadening function $\Theta(\lambda)$, and rotation function  $G(\lambda)$: 

\begin{equation}\label{Eq8}
  D(\lambda)= G(\lambda)\ast \Theta(\lambda)\ast I(\lambda)\ast F_{\lambda}^0.
\end{equation}
Here, an asterisk denotes convolution as a certain special operation. If one combines the rotation and macroturbulence functions into a single macrobroadening function $M(\lambda)=G(\lambda)\ast\Theta(\lambda)$, Eq.~(1) may be presented as a double convolution:

\[ D(\lambda)= M(\lambda)\ast I(\lambda)\ast F_{\lambda}^0.\] 
The convolution is replaced by a product in the Fourier transform domain, and the transform of the observed line profile is written as

\begin{equation}\label{Eq10}
  d(\sigma) =m(\sigma)i(\sigma)f_{\lambda}^0(\sigma),
\end{equation}
where lowercase letters denote transformations (or transforms) of the corresponding functions, and $\sigma$ is the Fourier frequency. If $\lambda$ is given in nanometers, $\sigma$ is expressed in cycles/nm (or nm$^{-1})$; if $\lambda$  is set in velocity units (km/s), $\sigma$ is expressed in s/km.

Since transforms $d(\sigma)$, $m(\sigma)$, $i(\sigma)$, and $f_{\lambda}^0(\sigma)$ are needed for Fourier analysis, we should first calculate the corresponding functions. Let us start with function $F_{\lambda}^0$ that characterizes the true (intrinsic) line flux profile of a nonrotating star without macroturbulenceprior to processing in the spectrograph. This function is given by  

\[F_{\lambda}^0 = 2\pi\int_0^{\pi/2}I_{\lambda}^0\sin \theta\cos \theta d\theta, \] 
where $I_{\lambda}^0$ is the profile of specific intensity without macroturbulence, which includes the components of  atomic, thermal, and microturbulent broadening; $\theta$ is the angle between the normal to the stellar surface  and the line of sight.  

Microturbulence is set by a Gaussian distribution function with variance  $\xi_{\rm  mic}$ and is introduced as a convolution with the thermal plus atomic line absorption coefficient at each depth within the atmosphere. In the general case, intrinsic profile $F_{\lambda}^0$ is calculated by solving the transfer equation for the given atmosphere model.

The motion of gas in the photosphere on scales exceeding the average free path length of a photon is factored in by introducing the macroturbulence function that may be written as an isotropic Gaussian function:

\begin{equation}\label{Eq3}
\Theta(\xi)=  \frac{1}{\xi_{\rm mac}\sqrt{\pi}}\exp\left(-\frac{\xi^2}{\xi_{\rm mac}^2}\right).
\end{equation}
Here, macroturbulent velocity $\xi_{\rm  mac}$, which does not vary with height in the atmosphere, is the variance.   Gray  \cite{1976oasp.book.....G, 1977ApJ...218..530G} has proposed to use a radial-tangential macroturbulent velocity distribution function in the form of a sum of two Gaussian functions under the assumption of purely radial and purely tangential motions with velocities  $\xi_{\rm R}$ and $\xi_{\rm T}$ occupying areas  $S_{\rm R}$ and $S_{\rm T}$ on the disk surface. Neglecting the variation of intensity from the center to the limb and using the radial-tangential macroturbulence function, one obtains a more complex form of Eq. (1):

\[ D(\lambda)= G(\lambda)\ast 2[S_{\rm R} \Theta(\xi_{\rm R}) +  S_{\rm T}\Theta(\xi_{\rm T})]\ast I(\lambda)\ast F^0_\lambda.\]
Here, radial  $\Theta(\xi_{\rm R})$ and tangential  $\Theta(\xi_{\rm T})$ functions are written as 

\begin{equation}\label{Eq0a}
  \Theta(\xi)= (\pi\xi)^{1/2}(\lambda/\xi)\int_0^{\xi/\lambda} \exp(-1/u^2)du,
\end{equation}
where $u=(\xi_{\rm R}\cos\theta)/\Delta\lambda$  for   $\Theta(\xi_{\rm R})$ and  $u=(\xi_{\rm T}\sin\theta)/\Delta\lambda$ for  $\Theta(\xi_{\rm T})$. 

The simple case with  $S_{\rm R}=S_{\rm T}=0.5$  and $\xi_{\rm R}=\xi_{\rm T}=\xi_{\rm RT}$ is often used for Fourier analysis.

The effect of rotation on the line profile is modeled by rotation function $G(\lambda)$ that is given by the well known analytical expression \cite{1973ApJ...184..461G, 1976oasp.book.....G}

\begin{equation}\label{Eq5}
G(x)=c_1(1-x^2)^{1/2}+c_2(1-x^2),
\end{equation}
where $c_1=2(1-\epsilon)$, $c_2=0.5\pi\epsilon$ and $\epsilon$ is the coefficient of disk darkening toward the limb. This coefficient is determined from the law of intensity darkening in a continuum spectrum:

 \begin{equation}\label{Eq6}
I_c(\theta )=I_c(0)(1-\epsilon+\epsilon \cos \theta).
\end{equation}

Parameter $\epsilon$  varies from one star to another and depends on  $\lambda$. Parameter  $x=\Delta \lambda c/(\lambda v \sin i)$ for $|x|\leq1$, where $i$ is the angle between the rotation axis and the line of sight, $v$ is the rotation velocity at the equator, and $v\sin i$ is the component of rotation velocity along the line of sight that induces Doppler shift  $\Delta \lambda$.  The other designations are commonly known.

The distortion and tailing of stellar spectra induced by an insufficient spectral resolution of a spectrograph and other effects are characterized by instrumental function $I(\lambda)$ that may be written as a Gaussian function:

\[I(\lambda)=\frac{1}{\beta\sqrt{\pi}}\exp\left(-\frac{\Delta \lambda}{\beta}\right)^2,\]
where variance  $\beta=\lambda V/c$, and $V=\rm{FWHM}/(2\sqrt{\ln2})$. The full width at half maximum of the instrumental profile FWHM$~= c/R$, where  $R$ is the resolving power of a spectrograph.

Our goal in the present study is to determine the parameters of microturbulent, macroturbulent, and rotation velocities (i.e., $\xi_{\rm  mic}$, $\xi_{\rm  mac}$, and  $v\sin i$, respectively) using  Eq.~(\ref{Eq10}). Fourier analysis generally provides an opportunity to determine $\xi_{\rm  mic}$ by examining the position of the first side lobe in the observed transform $d(\sigma){/}i(\sigma)$. Setting $\xi_{\rm  mic}$, we calculate the  $f_{\lambda}^0(\sigma)$  transform and find its best fit to  $d(\sigma){/}i(\sigma)$. The value of  $\xi_{\rm  mic}$ is thus obtained, and the intrinsic $f_{\lambda}^0(\sigma)$ profile becomes known. Residual transform $m(\sigma)= d(\sigma){/}i(\sigma)/f_{\lambda}^0(\sigma)$  is then determined from Eq.~(2). Further interpretation of residual transform $m(\sigma)$ allows one to determine the remaining free parameters  $\xi_{\rm  mac}$ and  $v\sin i$.  In order to do this, we calculate the product of transforms of rotation and macroturbulence functions and compare it to the $m(\sigma)$ transform. The best fit defines the unknown velocity parameters $\xi_{\rm  mac}$ and  $v\sin i$. The procedure was detailed in \cite{1977ApJ...218..530G}.

This Fourier method has an advantage in that it allows one to substitute a complex mathematical operation (convolution of functions) with simple multiplication or division of transforms. Its weak point is the assumption that the line intensity profile does not depend on the position at the disk.  

In view of this, Smith and Dominy \cite{1979ApJ...231..477S} have proposed to use an integral method, where a transform of the observed line profile is considered instead of the residual transform and is compared with the transform of the line profile calculated by numerical integration over the entire disk with the variation of line intensity from the disk center to the limb factored in Bruning  \cite{1984ApJ...281..830B}  has tested these two versions of the Fourier technique and concluded that the disregard of intensity variation from the center to the limb may result in a systematic overestimation (or underestimation) of line broadening, which depends on the line strength, in the case of late-type stars. Certain solar lines exhibit several-percent variation of $I_{\lambda}^0$  as a function of  $\theta$, while the intensities of other lines at the disk center and at the limb are almost equal. The majority of solar lines reveal relatively large variations only near the limb (i.e., within a small part of the disk area). Therefore, approximation $I_{\lambda}^0=$~const does not introduce any considerable errors. If variations from the center to the limb are weaker than broadening by the dominant mechanism (e.g., rotation for stars with  $v\sin i >10$ km/s, this approximation has no effect on the Fourier analysis results.

At the same time, the integral method \cite{1979ApJ...231..477S} is almost useless if all three parameters   $\xi_{\rm  mic}$, $\xi_{\rm  mac}$, and  $v\sin i$ are unknown and need to be determined. The reason for this is that only the transforms of observed and calculated line profiles are used in the process of fitting: the resulting uncertainty is much higher than the one obtained when the observed, intrinsic, and residual transforms are used.

\section{Specifics of Fourier analysis of stellar spectra }

The Nyquist frequency  $ \sigma_N$ is a parameter in Fourier analysis defined as the highest of the working frequencies in a transform that allow for correct recovery of the function. This frequency is written  $ \sigma_N = {0.5}/{\Delta \lambda} $, where  $\Delta \lambda$ is the sampling interval. If, for example, $\Delta \lambda =1$ pm, the Nyquist frequency is  $\sigma_N = 500$~cycles/nm. This means that all frequencies below  500 cycles/nm  may technically be analyzed if they are above the noise level.

Observation errors, which produce noticeable constant white noise in the observed profile transform, affect the accuracy of results of Fourier analysis. White noise depends on the line measurement errors  rather than on the line strength. According to  \cite{1976oasp.book.....G, 1976ApJ...208..487S}, ¾the white noise level in a transform may be determined using the following simple formula:

\[  S(\sigma) = S(\lambda) \Delta \lambda \sqrt N,\]
where $S(\lambda)$ is the observational data error (noise-to-signal ratio), $N$ is the number of points in the observed line profile, and $\Delta \lambda$ is the distance between the measured profile points. An increase in the signal-to-noise ratio in observations corresponds to a reduction in the noise level and provides an opportunity to analyze higher frequencies in the line transform. It is preferable for Fourier analysis to have a signal-to-noise ratio within the interval of  300--1000, but even this does not make the observed profiles free from blending by weak lines that introduce additional noise into the transform. According \cite{1976ApJ...207..308S}, if the blending line is weak, it manifests itself in the Fourier transform of a line as a small-amplitude sinusoidal filter. Therefore, it is important for Fourier analysis to obtain unblended and symmetrical line profiles with wings evolving smoothly into continuum. However, difficulties in the fitting of side lobes at high frequencies in residual transforms, which are not always that easy to interpret, may arise even when the noise level is relatively low, lines are unblended, and the resolution is high.

The stellar rotation velocity may be the cause of these difficulties. Smith  \cite{1979PASP...91..737S}  has paid particular attention to restrictions in the detection of small rotation velocities. The higher is the $v\sin i$, parameter, the more zeros and side lobes emerge in the rotation function transform, and the easier it is to distinguish between the effects of rotation and macroturbulence in the observed line profile transforms.  

In the case of slowly rotating stars, the rotation function transform has little influence on the observed transform. Side lobes are often not visible in the rotation function transform, and one is forced to work with just the main lobe. The effects of intrinsic, macroturbulent, and rotational line profile broadening become comparable. The combination of their transforms defines the working frequency region for analysis.
The rotational transform has a flat-topped shape in such cases, while the intrinsic profile and macroturbulence transforms drop rapidly at low frequencies. As a result, the observed line transform rapidly drops to the noise level, and it becomes difficult to measure the parameters of macrobroadening velocities, while the microturbulence velocity may be determined reliably. This means that even a high signal-tonoise
ratio and a fine resolution do not enable one to recover high frequencies and obtain reliable results for  $\xi_{\rm mac}$, $v \sin i$.

Smith \cite{1979PASP...91..737S} believed that   $v \sin i=2$--2.5 km/s is the lowest rotation velocity that may be distinguished reliably from macroturbulence and measured. He divided slowly rotating stars into the following three categories:

 (1) Stars with low rotation velocities (6--10 km/s) that may be measured using the position of the first zero and the side lobe shape. Their analysis may be complicated by deviations from local thermodynamic equilibrium (LTE) and blending in the continuum.

(2) Stars with very low rotation velocities (2.5--6 km/s) that may be measured only based on the shape of the main lobe with difficulties arising due to incomplete knowledge of the instrumental profile and uncertainties of microturbulence and macroturbulence parameters.

(3) Stars with ultralow rotation velocities ($< 2$--2.5~km/s) that set the ultimate limit of velocity determination by the discussed method.

Solar-type stars belong to the third group. Their spectra contain virtually no unblended lines, it is not always possible to determine the continuum reliably, and the effects of deviation from LTE may also manifest themselves. Therefore, Fourier analysis of these stars is especially challenging, since one is forced to analyze only the main lobe with the values of  $\xi_{\rm mic}$, $\xi_{\rm mac}$, and $v \sin i$ remaining unknown. The width of the frequency region suitable for analysis in the main lobe is relatively small and may vary from one star to another and even from line to line.

\section{Procedure of Fourier analysis for stars \\ with very slow rotation  }

We have tested the Fourier analysis techniques developed by Gray  \cite{1973ApJ...184..461G, 1975ApJ...202..148G, 1977ApJ...218..530G, 1982ApJ...262..682G, 2006PASP..118.1112G} and the integral
Fourier method proposed by Smith et al. \cite{1976ApJ...208..487S, 1979ApJ...231..477S, 1976ApJ...207..308S} by applying them to solar-type stars. It turned out that basically only the main lobe is analyzable, and the obtained results are mixed. It was concluded that the Fourier method based on the convolution theorem is more reliable in this case. We have failed to determine  $\xi_{\rm mic}$  based on the position of the first zero in transform  $d(\sigma)$ in the limiting case of slow stars ($v\sin i<2$~km/s). This complicated the task. Therefore, our technique differs from the ones developed earlier. Fourier transforms of all functions were calculated using the fast Fourier transform program written by K. Pikalov. Each step of the procedure used in the present study is detailed below.

Step 1. In accordance with the requirements of Fourier analysis, we express the spectral line profiles in terms of line depth so as to make the profile smoothly go to zero at both sides. The chosen line is examined for the presence of invisible blends that may induce asymmetry of the line profile. The line synthesis method, an atmosphere model, and a complete list of blends are used for this purpose. If unblended lines are not available, we choose a line with wings that are largely free from blends (or a line with at least one clean wing). The profile is then corrected in such a way as to make it symmetric and free from blends and make its wings go to continuum smoothly. In order to do that, we calculate the line profile using the photosphere model and fit the obtained result to clean sections of the observed profile. A single broadening parameter (or several broadening parameters) and element abundance  $A$. may be treated as free parameters at this stage. Equivalent width $W_{\rm obs}$ and Fourier transform $d(\sigma)$ are calculated for the corrected observed profile.

Step 2. Intrinsic line profile  $F_{\lambda}^0$  0 is calculated in the LTE approximation using the atmosphere model and an isotropic Gaussian distribution of microturbulent velocities for a set of  $\xi_{\rm mic}$ values. Equivalent width  $W$ of the intrinsic profile is brought into agreement with $W_{\rm obs}$ by varying abundance  $A$. The $W=W_{\rm obs}$ equality guarantees the equality of maximum amplitudes of the main lobe in transforms  $d(\sigma)/i(\sigma)$ and $f_{\lambda}^0(\sigma)$. This condition is essential in Fourier analysis and should be satisfied. Fourier transforms $f_{\lambda}^0(\sigma,\xi_{\rm mic},A$) are calculated for the obtained set of intrinsic profiles.

Step 3. The instrumental profile and its Fourier transform $i(\sigma)$.  are calculated. Resolution $R = 300 000$ and 48 000,  FWHM~$= c/R = 1$ and 6.25~km/s, and parameter $V=\rm{FWHM}/(2\sqrt{\ln2})=0.6$ and 3.75~km/s for the solar spectrum and stellar spectra, respectively.

Step 4. Macroturbulent broadening function  $\Theta(\lambda)$ is calculated using both isotropic distribution (\ref{Eq3}) with  $\xi_{\rm mac} =$~const and nonisotropic radial-tangential distribution (\ref{Eq0a}) with $\xi_{\rm RT} =$~const for all photosphere layers. Rotation function  $G(\lambda)$ from (\ref{Eq5}) is calculated with the coefficient of darkening at the limb that is determined in advance using Eq.~(\ref{Eq6}) and the atmosphere model. Setting the initial rotation velocity (e.g., $v \sin i=1$ km/s) and a set of macroturbulent velocity values, we calculate a set of functions  $\Theta(\lambda)$ and $G(\lambda)$ and products of their transforms. These calculations are then repeated for another $v \sin i$  value with the same set of macroturbulent velocities, etc. A set of transforms  $m(\sigma, v\sin i,  \xi_{\rm mac})$ is thus obtained.

Step 5. Residual transforms are calculated by dividing  $d(\sigma)/i(\sigma)$ by each transform from the $f_{\lambda}^0(\sigma,\xi_{\rm mic},A)$ set obtained in Step 2. The first residual transform is compared to macrobroadening function transforms $m(\sigma, v \sin i,  \xi_{\rm mac})$ obtained in Step 4. The best fit is determined by finding the minimum mean-square deviation $\chi^2$. This procedure is repeated for the next residual transform with a different $\xi_{\rm mic}$  value, etc. The minimum  $\chi^2$ value, which determines all the parameters of interest ($\xi_{\rm mic}$, $\xi_{\rm mac}$, and $v\sin i$), is chosen from the obtained set. This iteration process requires visual monitoring.

Step 6. In order to verify the correctness of the obtained result, the line profile is synthesized using the atmosphere model and the values of parameters $\xi_{\rm mic}$, $\xi_{\rm mac}$, $v\sin i$, and $A$ determined via Fourier analysis. A satisfactory agreement between the calculated and the observed line profiles is verified by the minimum  $\chi^2$ value. If the profiles do not match, the procedure is repeated. Each step is checked, and certain changes are made (e.g., the free parameter increment is altered). It may be necessary to correct the observed profile once again or even choose a different line if it found impossible to remove all blends.

%%%%%%%%%%%%%%%%%%%%%%%%%%%%%%%%
\section{Initial data selection and synthesis of spectra}
%%%%%%%%%%%%%%%%%%%%%%%%%%%%%%%%

We have chosen the lines with equivalent widths $W = 2$--20~pm (Table 1) from a comprehensive list of neutral iron lines. The combination and the number of chosen lines vary somewhat from one star to another due to differences in the line strength and blending. The complete list of blends in the wavelength range of chosen lines was taken from the atomic spectral line database compiled by Kurucz (CD-ROM 23, http://www.pmp.uni-hannover.de/cgi-bin/ssi/test/kurucz/sekur.html) and from the VALD database \cite{1999A&AS..138..119K}. The data for oscillator strengths $\log gf$ were taken from \cite{2006JPCRD..35.1669F}, where a compilation of experimental data was presented. We have used only the $\log gf$ values with uncertainties within $\pm 3$\% and $\pm 10$\%.

The spectrum of the Sun as a star was taken from   \cite{2005ASPC..336..321H}. Its spectral resolution is approximately 300000. The observed spectra of HD 1835 and HD 10700 and the fundamental parameters of these stars were provided by Ya. Pavlenko and A. Ivanyuk. The observations were carried out by Jenkins et al. \cite{2008A&A...485..571J} at the 2.2-meter MPG/ESO telescope in Chile with the FEROS spectrograph with a resolving power of 48000 and a sign-to-noise ratio exceeding 150.

%%============================================\lfloor==============

%___________________________________ Table 1
\begin{table}
\centering
 \caption{\small 
  Parameters of lines selected for Fourier analysis (lower excitation potential $E_{\rm exp}$, oscillator strengths $\log gf$, central line depths $D$, equivalent widths $W_{\lambda}$, and average effective formation depths $\log\tau_{5,W}$) and the obtained estimates of macroturbulent radial-tangential $\xi_{\rm RT}$ and isotropic $\xi_{\rm mac}$ velocities, microturbulent velocity $\xi_{\rm mic}$, rotation velocity $v\sin i$, and iron abundance $A$  } 
\vspace {0.3 cm}
\label{T:1}
\footnotesize{
%\small{
\begin{tabular}{ccccccccccc}
\hline\hline
 $\lambda$ & $E_{\rm exp}$   & $\log gf$&$D$& $W_{\lambda}$&$\log\tau_{5,W}$&$\xi_{\rm RT}$  &  $\xi_{\rm mac}$ & $\xi_{\rm mic}$& $v\sin i$  & $A$  \\
 (nm)       & (eV)     &          & &(pm)&   & (km/s )    & (km/s )   &  (km/s )   &  (km/s )    & \\
\hline
     \multicolumn{9}{c} {THE SUN IS A STAR}  \\
448.42249 & 3.603 &  -0.864  & 0.800 & 11.47& -1.48  &  2.95&   2.05   &   0.8 &  1.74 & 7.640\\
460.20044 & 1.608 &  -3.154  & 0.726 &  7.48& -1.88  &  3.05&   2.05   &   0.8 &  1.76 & 7.596\\
499.41363 & 2.198 &  -1.136  & 0.809 & 11.49& -2.42  &  2.9 &   2.0    &   0.9 &  1.80 & 7.514\\
504.98243 & 2.279 &  -1.355  & 0.823 & 17.27& -2.06  &  2.9 &   2.00   &   0.75 & 1.71 & 7.582\\
508.33395 & 0.958 &  -2.958  & 0.807 & 11.83& -2.46  &  2.9 &   2.0    &   0.9 &  1.71 & 7.482\\
524.24920 & 3.635 &  -0.967  & 0.699 &  9.44& -1.63  &  2.95&   2.05   &   0.7 &  1.77 & 7.635\\
524.70497 & 0.087 &  -4.946  & 0.660 &  6.89& -2.09  &  2.85&   1.95   &   0.8 &  1.79 & 7.622\\
525.02085 & 0.121 &  -4.938  & 0.651 &  6.91& -2.08  &  3.0 &   2.1    &   0.8 &  1.72 & 7.658\\
549.75146 & 1.011 &  -2.849  & 0.786 & 12.90& -2.50  &  2.7 &   1.85   &   0.9 &  1.81 & 7.492\\
550.14605 & 0.958 &  -3.047  & 0.774 & 12.09& -2.49  &  2.85&   1.95   &   0.9 &  1.74 & 7.521\\
550.67793 & 0.990 &  -2.797  & 0.794 & 13.31& -2.57  &  2.65&   1.85   &   0.95&  1.75 & 7.464\\
566.13418 & 4.285 &  -1.756  & 0.212 &  2.24& -0.64  &  3.05&   2.1    &   1.0 &  1.72 & 7.391\\
570.54616 & 4.302 &  -1.355  & 0.344 &  3.96& -0.81  &  3.3 &   2.3    &   0.8 &  1.68 & 7.408\\
577.84518 & 2.588 &  -3.430  & 0.214 &  2.28& -0.97  &  3.3 &   2.25   &   0.9 &  1.77 & 7.440\\
   \multicolumn{9}{c} {HD 10700}  \\
460.19947 & 1.608  & -3.154 &0.508&   7.27&  -1.77&   2.4   &  1.7    &   0.4 & 0.79 &  7.140\\
499.41265 & 2.198  & -1.136 &0.648&  12.07&  -2.24&   0.55  &  0.4    &   0.5 & 0.79 &  7.025\\
524.24940 & 3.635  & -0.967 &0.480&   8.17&  -1.43&   2.1   &  1.5    &   0.7 & 0.78 &  6.990\\
537.95734 & 3.695  & -1.514 &0.345&   5.27&  -1.13&   2.4   &  1.7    &   0.4 & 0.80 &  7.090\\
541.27820 & 4.435  & -1.716 &0.085&   1.32&  -0.50&   2.3   &  1.6    &   0.8 & 0.78 &  6.990\\
549.75121 & 1.011  & -2.849 &0.650&  14.57&  -2.26&   0.85  &  0.6    &   0.7 & 0.78 &  7.031\\
550.14629 & 0.958  & -3.047 &0.622&  13.10&  -2.28&   1.15  &  0.80   &   0.7 & 0.78 &  7.032\\
550.67764 & 0.990  & -2.797 &0.649&  14.44&  -2.28&   0.40  &  0.30   &   0.7 & 0.75 &  6.945\\
566.13464 & 4.285  & -1.756 &0.092&   1.48&  -0.54&   2.60  &  1.85   &   0.4 & 0.80 &  6.940\\
577.84545 & 2.588  & -3.430 &0.121&   1.94&  -0.91&   3.05  &  2.05   &   0.5 & 0.78 &  7.016\\
\multicolumn{9}{c} {HD 1835}  \\
448.42307 & 3.603&  -0.864  &0.568&  13.70& -1.27&  5.0   &  3.4   &   1.3   & 6.21  &  7.822 \\
460.20048 & 1.608&  -3.154  &0.441&   9.22& -1.66&  5.25  &  3.65  &   1.2   & 6.26  &  7.818 \\
524.24919 & 3.635&  -0.967  &0.454&  11.64& -1.33&  4.85  &  3.55  &   1.1   & 6.26  &  7.890 \\
537.95767 & 3.695&  -1.514  &0.323&   7.86& -0.88&  5.45  &  3.75  &   1.0   & 6.29  &  7.868 \\
549.75028 & 1.011&  -2.849  &0.579&  17.02& -2.30&  5.1   &  3.4   &   1.25  & 6.23  &  7.858 \\
550.14664 & 0.958&  -3.047  &0.535&  15.16& -2.33&  5.0   &  3.4   &   1.2   & 6.21  &  7.839 \\
550.67787 & 0.990&  -2.797  &0.560&  17.09& -2.29&  5.45  &  3.8   &   1.1   & 6.24  &  7.841 \\
\hline

\end{tabular}
}
\end{table}
\noindent
%%%%%%%%%%%%%%%%%%%%%%%%%%%%%%%%%%\lhd%%%%%%%%%%%%%%%%%%

%______________________________________________________

%%==========================================================

%___________________________________ Table 2
\begin{table}
\centering
 \caption{\small 
 Parameters of stars (effective temperature $T_{\rm eff}$, surface gravity $\log g$, and metallicity [M/H]) and the average values of velocities and iron abundance $A$ obtained using Fourier analysis.  The upper $\xi_{\rm RT}$ and $\xi_{\rm mac}$ estimates were obtained based on weak lines; the lower ones are based on stronger lines.
 } \vspace {0.3 cm}
\label{T:2}

\footnotesize
\begin{tabular}{ccccccccc}
\hline\hline
 Star&$T_{\rm eff}$&$\log g$&[M/H]&$\xi_{\rm RT}$& $\xi_{\rm mac}$&$\xi_{\rm mic}$& $v\sin i$& $A$  \\
        & (K)     &          &    & (km/s)    & (km/s)   &  (km/s)   &  (km/s)    & \\
\hline
 Sun  &   5777&4.44& 0.00& $3.22\pm0.14$& $2.22\pm0.10$& $0.85\pm0.09$& $1.75\pm0.04$&$7.53\pm0.09$\\
         &       &    &     & $2.89\pm0.12$& $1.99\pm0.08$&                &                &                \\
 HD 10700 &   5383&4.59&-0.60& $2.47\pm0.32$& $1.73\pm0.19$& $0.58\pm0.15$& $0.78\pm0.01$&$7.02\pm0.06$\\
        &       &    &      & $0.74\pm0.33$& $0.52\pm0.22$&                &                &                \\
 HD 1835  &   5807&4.47& 0.20& $5.15\pm0.23$& $3.56\pm0.17$& $1.16\pm0.10$& $6.24\pm0.03$&$7.85\pm0.03$\\
        \hline
\end{tabular}
\end{table}
\noindent

Spectral lines were synthesized using the SPANSAT code  \cite{1988ITF...87P....3G}  with the LTE condition assumed in all calculations. The Gaussian distribution of microturbulent velocities is factored in by convolving with the line absorption coefficient in each atmospheric layer. Macroturbulent motions were taken into account using a macroturbulent velocity field with either an isotropic Gaussian distribution or a radial-tangential distribution. The stellar rotation was factored in by direct integration of intensity over the disk. The van der Waals damping constant was calculated according to the Anstee-Barklem-O'Mara method. The needed damping parameters s and a were taken from  {\cite{2005A&A...435..373B, 2000A&AS..142..467B}. The instrumental profile was assumed to be Gaussian and is factored in by convolving with the line profile.

The solar atmosphere model was taken from the MARCS database  \cite{2008A&A...486..951G}. Atmosphere models for the stars were obtained by interpolating the MARCS data for the following fundamental parameters adopted in the present study: effective temperature  $ T_{\rm eff}$, surface gravity $ \log g$, and metallicity [M/H] (Table 2). The chemical composition of the Sun corresponds to the data presented by Asplund et al.  \cite{2005ASPC..336...25A}.  The law of ideal gas composed of molecules, atoms, and ions was used in the recalculation of atmosphere models. Radiation pressure was also taken into account. The atmospheric opacity was calculated using the Kurucz programs \cite{1970SAOSR.309.....K} with a wide range of absorbers.

\section{Results and discussion }

\subsection{The Sun as a Star }

The Sun had the highest number of selected lines (14). A typical example of an observed profile after correction and symmetrization is shown in Fig. 1a, and the Fourier transform of this profile and transforms of other functions are presented in Fig. 1b. The best fit of the residual transform of the same line to the macrobroadening function was obtained in the main lobe region in the frequency interval extending to  $\sigma = 60$~nm$^{-1}$. The discrepancies start growing larger in the region where the intrinsic transform decays rapidly. Although this line with  $W= 12.1$ pm has a well-marked side lobe and the first zero near $\sigma = 95$~nm$^{-1}$, is seen clearly, it is impossible to align them. They are located near the noise level (approximately $-3.5$). If $\xi_{\rm mic}$ and $A$ are chosen so that their positions match, a poor fit in the main lobe region is obtained. The reference line profile calculated with parameters $\xi_ {\rm mic}$, $\xi_{\rm RT}$, $v \sin i$, and $A$ derived by matching the Fourier transforms is the criterion here. It turned out that the best match between the observed profile and the reference one is obtained when only the main lobe is fitted. Therefore, it may be concluded that an agreement both in the maximum amplitude and the major part of the main lobe in residual transforms is necessary.

The comparison of reference and observed profiles demonstrated that the central part of the observed profile is 0.5--2\% deeper in the best-fit scenario (Fig.~1a). This effect is observed for the majority of solar lines from our list (the weakest lines are an exception here). It is fair to assume that this is caused by neglect of center-to-limb variation of the intrinsic profile. An attempt to find a better fit by comparing flux profiles instead of transforms was unsuccessful. The effect of mismatch in the central part of the profile in moderately strong lines has been noted long ago in the studies into solar spectrum synthesis. Gehren \cite{2001A&A...366..981G} attributed this to the use of micro- and macroturbulence parameters that do not vary with height in the photosphere: in his view, such parameters do not allow one to reconstruct both wings and cores of moderately strong observed lines. The results of synthesis of solar lines with 3D models without the micro- and macroturbulence approximation \cite{2015A&A...573A..26S}) have revealed a similar mismatch, which was attributed to non-LTE effects that were not taken into account in the synthesis of lines. Apparently, this conclusion is correct, since the line core at LTE is less deep. Gray \cite{2014AJ....147...81G} has also noted this effect in the profiles of hotter stars and proposed to compensate the mismatch between the cores of synthesized and observed profiles by reducing the temperature in upper photospheric layers.

%%%%%%%%%%%%%%%%%%%%%%%%%%%%%%%%%%%%%%%%% Figure 1
 \begin{figure}
 \centerline{ \includegraphics    [scale=1. ]{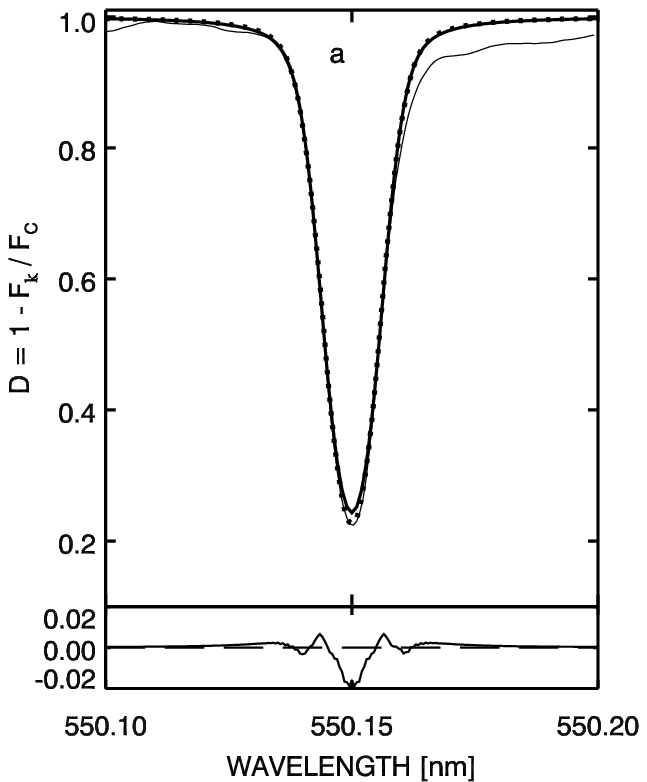}
 \includegraphics    [scale=1.]{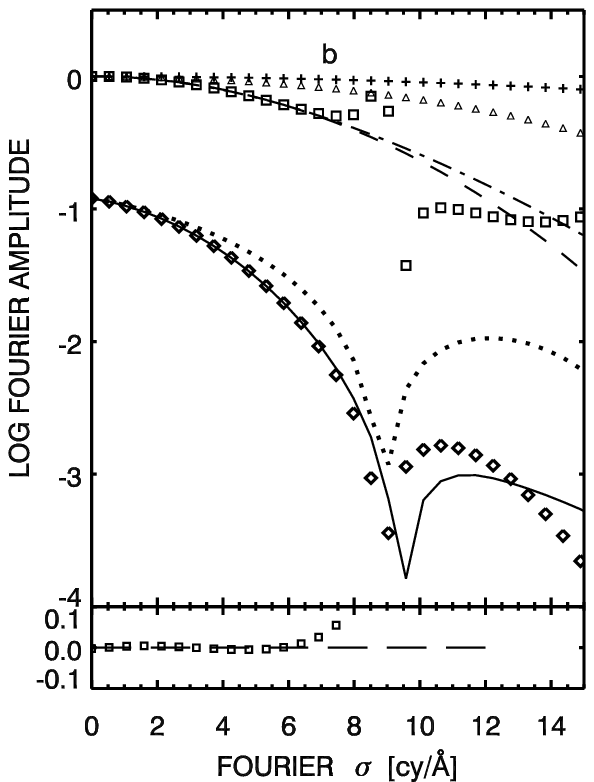}}
 \caption {\small
Sun: (a) observed (thin curve), corrected (dotted curve), reference (thick curve) profiles of a typical line; (b) Fourier transforms of the following profiles: observed (solid curve), intrinsic (dotted curve), combined isotropic macroturbulent with rotation (dashed curve), combined radial-tangential macroturbulent with rotation (dash-and-dot curve), rotation (triangles), instrumental (plus signs), residual (squares), and reconstructed (diamonds). Here, $\log A$ and $\sigma$ denote Fourier amplitude and Fourier frequency. The differences between calculated and observational data are shown at the bottom.
 } \label{abund2}
 %}
 \end{figure}
%%%%%%%%%%%%%%%%%%%%%%%%%%%%%%%%%%%%%%%%%%%%%%%%%%%%%%%%%%%%%
%%%%%%%%%%%%%%%%%%%%%%%%%%%%%%%%%%%%%%%%% Figure 2
 \begin{figure}
 \centerline{ \includegraphics    [scale=0.9 ]{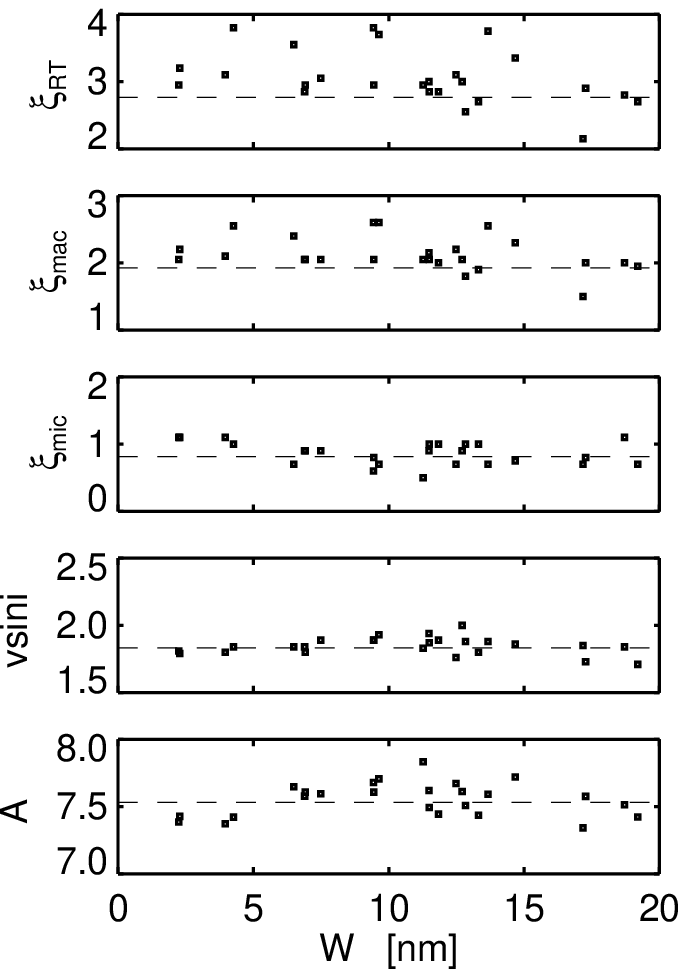}
 \includegraphics    [scale=0.9 ]{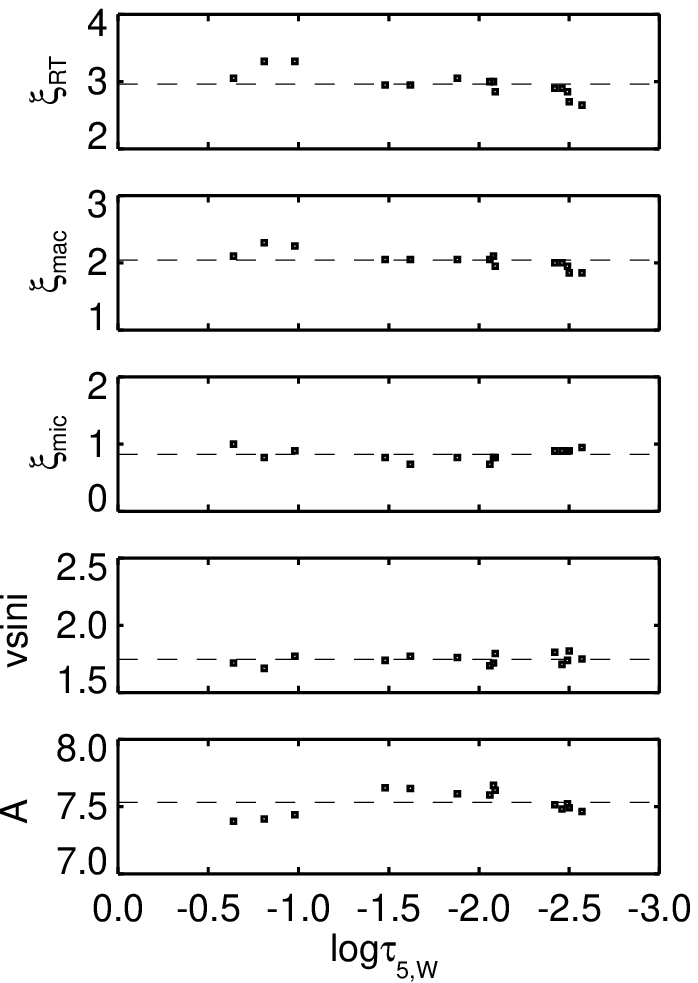}}
 \caption {\small
Variation of solar atmosphere parameters that were obtained by applying the Fourier analysis technique to each  line with (a) equivalent width and (b) average optical effective depth of line formation. The average values are represented by dashed lines. The velocities are given in km/s.
 } \label{abund2}
 %}
 \end{figure}
 %%%%%%%%%%%%%%%%%%%%%%%%%%%%%%%%%%%%%%%%%%%%%%%%
The velocities and the iron abundance estimated via Fourier analysis based on solar lines are presented  in Table 1 and in Fig. 2a. The obtained values of parameters  $\xi_{\rm RT}$ and $\xi_{\rm mac}$  decrease with an increase in the line strength. Weak lines form in the photosphere at a greater depth than strong ones and demonstrate that the velocities in the lower photosphere are much higher than those in the upper photosphere. In order to verify that, we have calculated the average optical effective depth of formation of the flux profile for each line:

 \[ \tau_{5,W} =\frac{1}{W}\int_{\lambda_1}^{\lambda_2} \tau_5(\lambda) D(\lambda) d\lambda,\]
where

\[ \tau_5(\lambda) =\int_0^1 \tau_{5}^\prime(\lambda,\mu) \mu d\mu\]
Ó $\tau_5^\prime(\lambda,\mu)$ is the optical effective depth of formation of a specific profile point in the position  $\mu= \cos \theta$ on
the stellar disk, and  $D(\lambda) $ is the line depth at the $\lambda$ profile point. $\tau_5^\prime(\lambda,\mu)$ is calculated as a weighted average over the depression contribution function  $F_D$. The calculation of $F_D$ was detailed in  \cite{1997MAO.....1P...3G}.

The obtained dependences of parameters on  $\tau_{5, W}$ (Fig. 2b) confirm the already known fact that the macroturbulent velocity tends to decrease with height in the photosphere. For example, it was demonstrated by Kostik  \cite{1982SoPh...78...39K},  Sheminova  \cite{1985KFNT....1...50S}, Gurtovenko and Sheminova \cite{1986SoPh..106..237G} that the macroturbulent velocity decreases with height due to the decay of convective motions and oscillations in the upper layers of the solar photosphere. It follows that one should not use a single averaged velocity for lines with different strengths in the synthesis of line profiles for flux. In order to illustrate the height dependence of macroturbulence, we have divided all lines into two groups. The first group is formed by lines with  $\log \tau_{5, W} >-1 $, which corresponds roughly to $W<5$ pm, and the second group contains all the remaining lines. The values of $\xi_{\rm mac}$ determined based on weak and stronger lines are 2.22 and 1.99 km/s. These results are in satisfactory agreement with literature data. For example, the values of $\xi_{\rm mac}=2.3$--1.9~km/s were obtained for the Sun as a star in \cite{1990SvA....34..260G, 1984BSolD1984...70S, 1998KPCB...14..169S, 1996A&AS..118..595V}, and $\xi_{\rm mac}=1.9$~km/s was found in \cite{1976SoPh...46...29C} by analyzing the data for the center of the solar disk and the limb after conversion to the full disk. The theoretical calculations in \cite{2013MSAIS..24...37S} based on 3D atmosphere models yielded $\xi_{\rm mac}=2.2$ km/s.

Let us focus our attention on the macroturbulence velocity in the radial-tangential model (RT model) approximation. The application of the RT model was discussed in \cite{1975ApJ...202..148G, 1976oasp.book.....G, 1977ApJ...218..530G, 1982ApJ...262..682G}, and it was concluded that this model reconstructs well the Fourier transform of observed profiles for late-type giant stars. It was suggested that the RT model should be suitable for the majority of stars in the H-R diagram. In the present analysis, the RT model was also used for all stars. Without going into detail, we may note that the RT model does differ from the common isotropic Gaussian model, and the RT model indeed yields a better fit of residual transforms only for stars with substantial macrobroadening. This difference is evident at higher frequencies. This is seen clearly in Fig. 1b, where the dashed curve represents the residual transform for the Gaussian model, and the dash-and-dot curve corresponds to the RT model. Since primarily the main lobe is used in the analysis of solar-type stars, the difference between two macroturbulence models
is very small. The obtained dependences for $\xi_{\rm RT}$ and $\xi_{\rm mac}$  are similar (Fig. 2b). The average results for weak and stronger lines ($\xi_{\rm RT}=3.22$ and 2.89~km/s) are in a satisfactory agreement with the results of other studies (e.g., 3.8--3.1 {\cite{1977ApJ...218..530G}},   $2.6$ {\cite{1998KPCB...14..169S}}, 4.0--2.3 {\cite{1995PASJ...47..337T}}, 3.2 {\cite{2001A&A...366..981G}}, 3.8--2.6 {\cite{2011A&A...528A..87M}}, 3.45~km/s {\cite{2013MSAIS..24...37S}}. It may be noted that the values obtained in different studies are scattered within the interval of  2.6--4~km/s.  This underscores the fact that lines of different strengths formed at different heights were used in these studies.

The results for other parameters, which do not vary with height, were averaged over all lines. The microturbulent velocity shows hardly any variation (within the limits of error) with $W$ and is virtually independent of the height in the photosphere (Fig.~2). Its average value is $\xi_{\rm mic}=0.85\pm0.09$~km/s. A velocity  of 0.8~km/s was obtained earlier for the Sun as a star in \cite{1998KPCB...14..169S}, where the Fourier method was also used. Other methods yielded a value of 0.75~km/s  \cite{2012MNRAS.422..542P, 1996A&AS..118..595V}.  Our result agrees well (within the accuracy of analysis) with the data obtained earlier.

%%%%%%%%%%%%%%%%%%%%%%%%%%%%%%%%%%%%%%%%% Figure 3
\begin{figure}
 \centerline{ \includegraphics    [scale=1.]{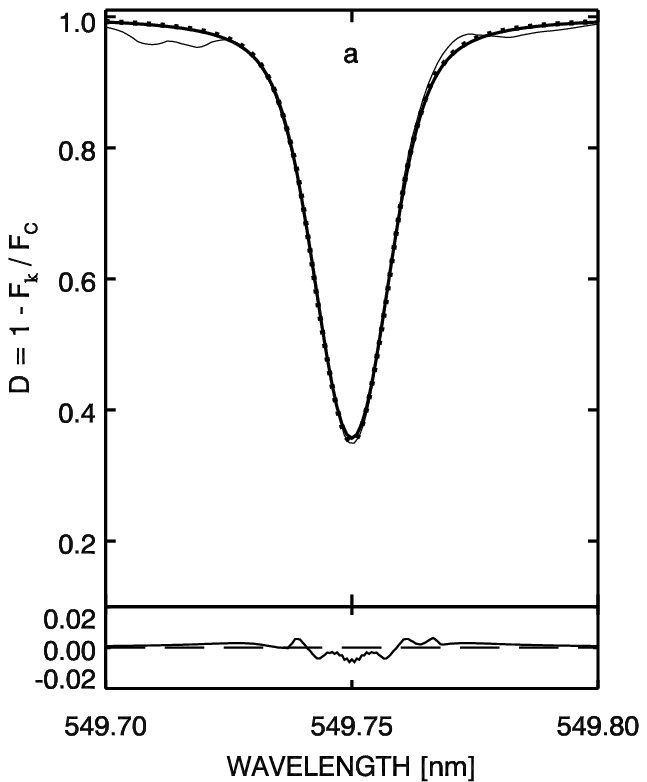}
 \includegraphics    [scale=1. ]{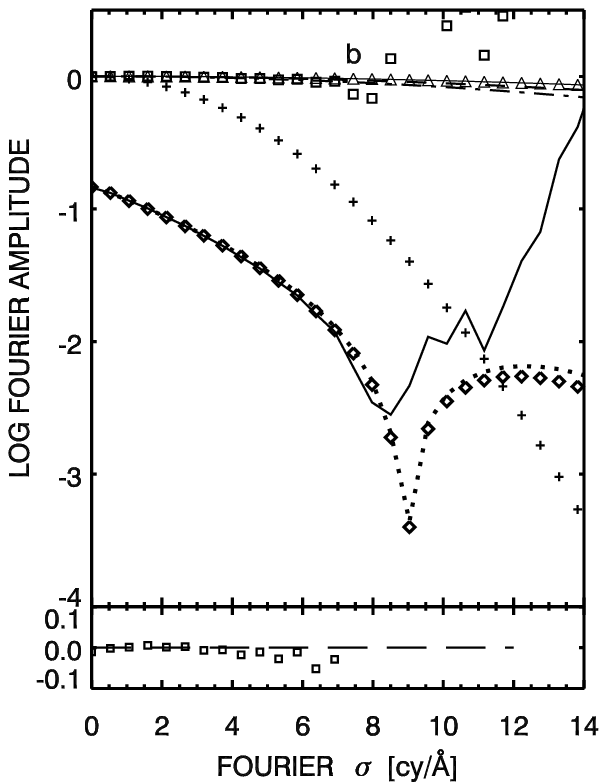}}
 \caption {\small
 Same as in Fig. 1 but for HD 10700.
 } \label{abund2}
 %}
 \end{figure}
%%%%%%%%%%%%%%%%%%%%%%%%%%%%%%%%%%%%%%%%% Figure 3
%%%%%%%%%%%%%%%%%%%%%%%%%%%%%%%%%%%%%%%%% Figure 4
\begin{figure}
 \centerline{ \includegraphics    [scale=1.0 ]{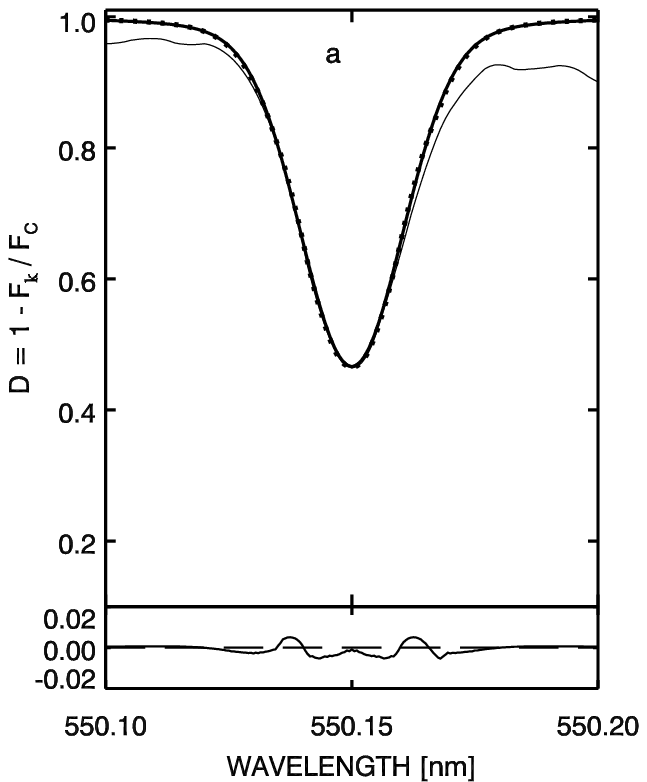}
 \includegraphics    [scale=1.0 ]{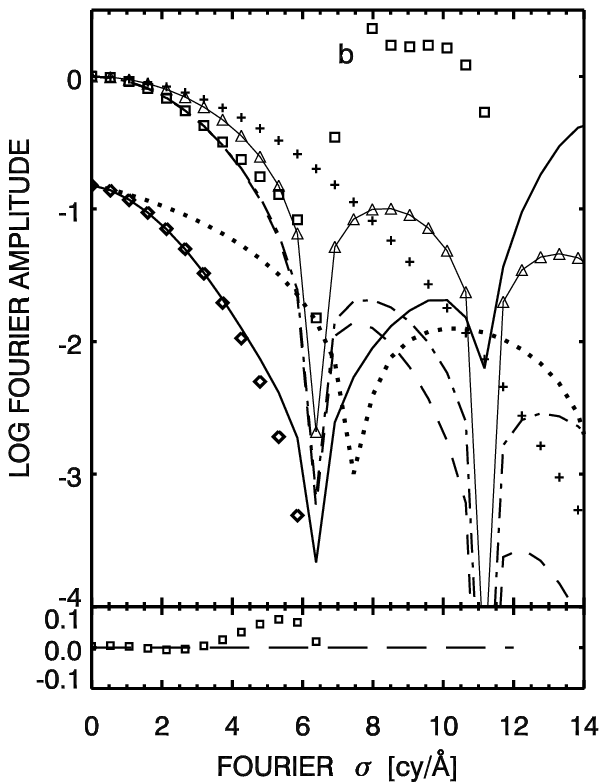}}
 \caption {\small
Same as in Fig. 1 but for HD 1835.
 } \label{abund2}
 %}
 \end{figure}
%%%%%%%%%%%%%%%%%%%%%%%%%%%%%%%%%%%%%%%%% Figure 4
The average rotation velocity determined in the present study is   $v \sin i =1.75\pm 0.04$~km/s. The obtained results confirm that the best fit of residual transforms is less sensitive to the variation of rotation velocity and macroturbulence than to the microturbulence velocity. It may be noted that the minimum deviation of residual transforms  $\chi^2$ changes little if $v \sin i$ is varied within the interval of  1.65--1.85~km/s. The same difference between results was found in other studies. A value of $v \sin i=1.6$~km/s was obtained in \cite{2012MNRAS.422..542P, 1996A&AS..118..595V} on line profiles, and  $v \sin i=1.7$~km/s was determined in  \cite{2005ApJS..159..141V}. An estimate of $1.85\pm0.1$~km/s was obtained in  \cite{1984ApJ...281..830B}  by Fourier analysis.

The determination of iron abundance A was not among the goals of the present analysis, but this parameter is adjusted together with the microturbulent velocity to match the equivalent width of the observed line profile. Provided that the oscillator strengths are reliable and the blend correction was done right, the value of  $A=7.53\pm0.09$ obtained in the present analysis should correspond to the commonly accepted value of $7.47\pm0.04$, which was calculated in  \cite{2015A&A...573A..26S} based on profiles at the solar disk center. Note also that Valenti and Piskunov \cite{1996A&AS..118..595V} have obtained $A=  7.49$ for the Sun as a star, and Pavlenko et al. \cite{2012MNRAS.422..542P} have reported $A= 7.56$.   A satisfactory fit between the iron abundance determined in the present study and the values found in other studies using different methods again confirms the validity of our results.

Thus, it follows from the comparison of results of Fourier analysis for the Sun with other data that this analysis method is suitable for estimating the macroturbulent velocity in solar-type stars if the microturbulent velocity and the rotation velocity are not known.

\subsection{HD 10700 and HD 1835}

The spectra of these two stars have lower resolution ($R =48 000$), lower signal-to-noise ratio (100--200), and, consequently, greater instrumental broadening of the line profile. The white noise level in the observed line profiles is approximately $-2.0$ (in logarithmic units).

Ten lines were analyzed for HD 10700. The position of the first zero and the side lobe in the transforms of observed lines are distorted by noise and are not analyzable. The line with $\lambda= 549.7$~nm and $W=14.6$ pm  is a fine example of that (Fig. 3). The intrinsic profile transform (dotted curve) almost coincides with the reconstructed observed line transform (diamonds). This means that the macrobroadening function has little effect on the observed profile, and its shape is defined solely by the microturbulent velocity. However, we have tried to derive the velocity parameters using the Fourier analysis technique.  

The obtained results are presented in Table 1 and in Fig. 5a for different values of  $\log\tau_{5, W}$.  Macroturbulent velocities in layers above $\log\tau_{5, W}=-2$ drop rapidly almost to zero.  If one divides all lines into two groups with the boundary between them set at  $\log\tau_{5, W}=-2$ or $W=8$ pm, and stronger lines yield the following estimates: $\xi_{\rm RT}=2.47\pm0.32$ and $0.74\pm0.33$~km/s;  $\xi_{\rm mac}=1.73\pm0.19$  and $0.52\pm0.22$~km/s, respectively. The microturbulent velocity averaged over all lines is  $0.58\pm0.15$~km/s, and rotation velocity  $ v \sin i  =   0.78\pm0.01$~km/s. The rotation velocity has little effect on the shape of the main lobe of this
star, since the rotation function transform is flat at low frequencies where the residual transforms are fitted. The $\chi^2$ minimum varies weakly in the  $ v \sin i  = 0.5$--1.5 km/s interval upon fitting the main lobe of the residual transform to the set macrobroadening function transform. The reference profiles calculated for each line confirm the smallness of  $ v \sin i$ (values (Fig. 3a).

Seven lines were processed for HD 1835. Since the Fourier transform of observed lines is filtered strongly by the rotation function profile, the position of the first zero in the line transform is governed by the rotation velocity, which is seen clearly at low frequencies.  Figure 4 shows the profile of the line with $\lambda= 550.1$~nm and $W=15.2$ mA pm and the corresponding Fourier transforms. It is seen clearly how the rotation function transform shapes the observed line transform. The positions of zeros in the rotation function and the observed line transform agree fairly well. The intrinsic profile affects only weakly the observed line shape.

%%%%%%%%%%%%%%%%%%%%%%%%%%%%%%%%%%%%%%%%% Figure 5
 \begin{figure}
 \centerline{ \includegraphics    [scale=1.0 ]{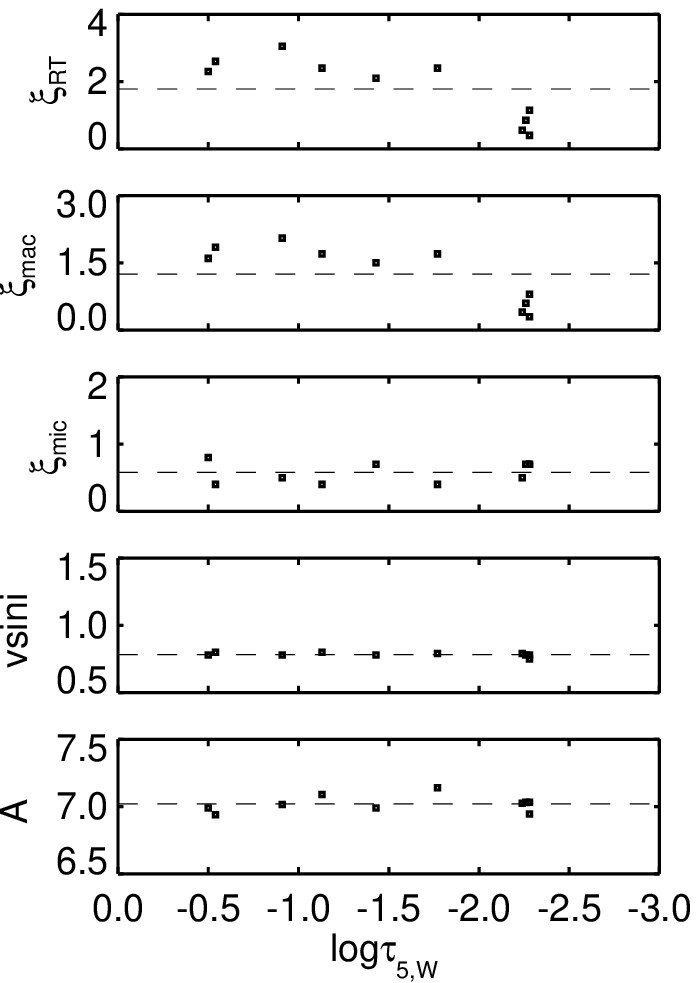}
 \includegraphics    [scale=1.0 ]{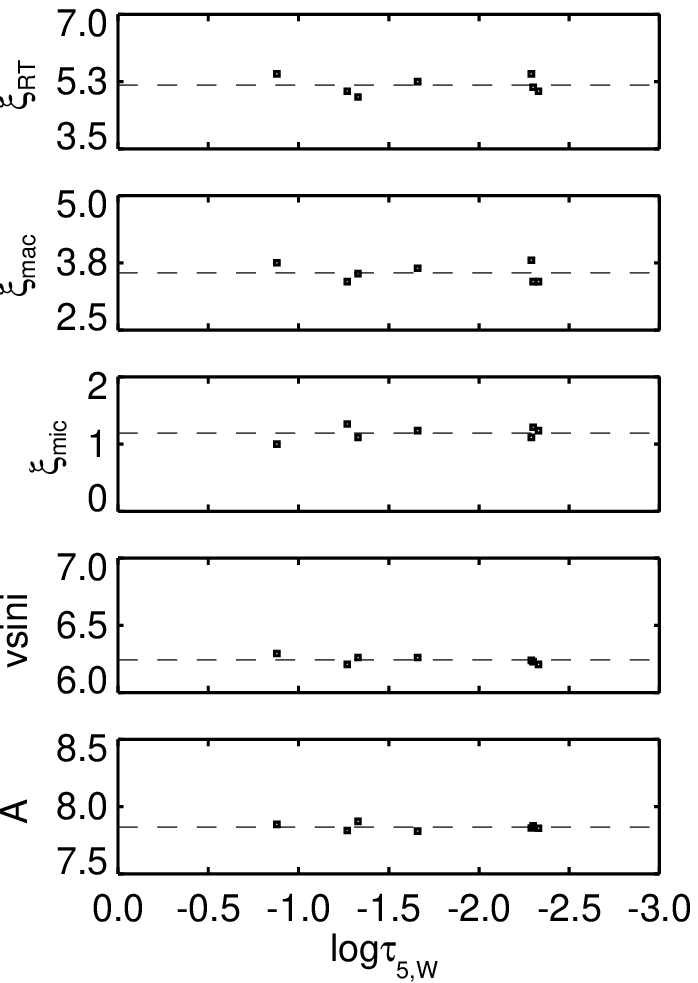}}
 \caption {\small
Variation of atmosphere parameters that were obtained by applying the Fourier analysis technique to each line with average effective depth of line formation for HD 10700 (left) and HD 1835 (right). The average values are represented by dashed lines. The velocities are given in km/s.
 } \label{abund2}
 \end{figure}
%%%%%%%%%%%%%%%%%%%%%%%%%%%%%%%%%%%%%%%%%%%%%%%%%%%%%%%%%%%%%%%%%%%%%

The obtained parameters for all lines are presented in Table 1 and Fig. 5b. Macroturbulent velocities reveal no apparent variation with height in the region of formation of the discussed lines. The results averaged over all lines are as follows:  $\xi_{\rm RT}=5.16\pm0.23$ and $\xi_{\rm mac}=3.56\pm0.17$ km/s. The average microturbulent velocity is $1.16\pm0.10$, , and the rotation velocity is $6.24\pm0.03$ km/s. The calculated reference profiles agree well with the observed ones (Fig. 4a), thus confirming the validity of the obtained results.

Let us compare our estimates (in brackets) with independent estimates obtained directly for these spectra. Pavlenko et al. \cite{2012MNRAS.422..542P} have reported  $\xi_{\rm mic}=0.5$~km/s  (0.58), $\xi_{\rm mac}  = 2.2 $~km/s  (1.73), and $v \sin i = 2.4\pm0.4$~km/s   (0.78) for HD 10700 and   $\xi_{\rm mic}=0.75$~km/s (1.16), $\xi_{\rm mac}  = 2.2 $~km/s (3.56), and $v \sin i = 7.2\pm0.5$~km/s (6.24) for HD 1835. It can be seen that the estimates differ. This may be associated with the fact that the macroturbulence velocity was not determined in  \cite{2012MNRAS.422..542P} and remained fixed. The estimated metallicity for HD 10700 and HD 1835 is [Fe/H]~$ = -0.38 \pm 0.04$ and $0.21\pm0.03$ (Jenkins et al., \cite{2008A&A...485..571J},  $-0.55$ and 0.23 (Pavlenko et al., \cite{2012MNRAS.422..542P}), and $-0.51$ and 0.32 (present study). Note that estimates obtained for these stars in several other studies may also be found in \cite{2012MNRAS.422..542P}.

\section{Conclusions}

The technique of application of Fourier analysis to spectral lines in real stellar spectra was developed by D.F. Gray. His book includes a very informative introduction into Fourier transform theory for beginners. In our opinion, the primary advantage of Fourier analysis over the classical method of synthesis with line profile fitting is that it provides an opportunity to distinguish between the effects of macroturbulence and rotation. This allows one to determine the macroturbulence parameter, which is no less important and significant in spectral analysis of stars than the microturbulence parameter. It should be mentioned that Fourier analysis defies automation, since it requires constant visual monitoring. It is an art that should be perfected.

We have tried to perform Fourier analysis for stars with ultraslow rotation. This is a limiting case. If the rotation velocity of a star is not known, Fourier analysis is the only way to estimate the macroturbulent velocity. Our tests demonstrated that the profile matching method produces less reliable results than Fourier analysis. This is the reason why we have chosen the Fourier technique. Its reliability and accuracy was demonstrated by applying Fourier analysis to the Sun as a star and two other solar-type stars with rotation velocities ranging from 1.0 to 7.0 km/s and comparing the obtained results with data from other studies.

We have also verified the applicability of Fourier analysis to stellar spectra with a low resolution of 48000. Low resolution, the lack of unblended line profiles, and the use of the LTE approximation limit the range of useful frequencies, thus complicating the process of fitting the residual transforms of calculated and observed profiles. In order to obtain reliable results under these conditions, one should use as many spectral lines as possible and perform Fourier analysis for each line separately. Our experience has shown that it is impractical to average out the residual transforms for lines of different strengths and then perform fitting for an averaged residual transform, since weak and strong lines may have different macroturbulence parameters. The lines that are affected weakly by damping and have the highest sensitivity to velocities and an equivalent width of 10--20 pm are the ones best suited for Fourier analysis. In addition, one may use weaker lines with $W$ as low as 2 pm. Very strong lines ($W > 20$~pm) with extended wings are not suitable for Fourier analysis, since they are defined by damping. Therefore, we recommend choosing moderately weak and moderately strong lines. These should first be carefully corrected for blends, which are almost always present, and only then subjected to Fourier analysis.

\vspace{0.3cm}
{\bf Acknowledgments.} I thank Ya. Pavlenko and A. Ivanyuk for providing the observed stellar spectra and the fundamental parameters of stars.

\end{document}